\documentclass[sigconf]{acmart}
\AtBeginDocument{%
  }

\setcopyright{cc}
\copyrightyear{2026}
\acmYear{2026}
\acmConference[TfT Workshop at CHI'26]{CHI'26 Workshop on Tools for Thought}{April 13--17, 2026}{Barcelona, Spain}
\acmISBN{978-1-4503-XXXX-X/2018/06}

\begin{document}

\title[Supporting Reasoning with Transparently Designed AI Data Science Processes]{More Than ``Means to an End'': Supporting Reasoning with Transparently Designed AI Data Science Processes}

\author{Venkatesh Sivaraman}
\email{venkatesh.sivaraman@ucsf.edu}
\orcid{0000-0002-6965-3961}

\author{Patrick Vossler}
\email{patrick.vossler@ucsf.edu}
\affiliation{%
\department{Weill Cancer Hub West}
  \institution{UC San Francisco}
  \city{San Francisco}
  \state{CA}
  \country{USA}
}

\author{Adam Perer}
\email{adamperer@cmu.edu}
\affiliation{%
  \institution{Carnegie Mellon University}
  \city{Pittsburgh}
  \state{PA}
  \country{USA}
}

\author{Julian Hong}
\email{julian.hong@ucsf.edu}

\author{Jean Feng}
\email{jean.feng@ucsf.edu}
\affiliation{%
  \department{Weill Cancer Hub West}
  \institution{UC San Francisco}
  \city{San Francisco}
  \state{CA}
  \country{USA}
}

\renewcommand{\shortauthors}{Sivaraman et al.}

\begin{abstract}
  Generative artificial intelligence (AI) tools can now help people perform complex data science tasks regardless of their expertise. While these tools have great potential to help more people work with data, their end-to-end approach does not support users in evaluating alternative approaches and reformulating problems, both critical to solving open-ended tasks in high-stakes domains. In this paper, we reflect on two AI data science systems designed for the medical setting and how they function as tools for thought. We find that success in these systems was driven by constructing AI workflows around intentionally-designed \textit{intermediate artifacts}, such as readable query languages, concept definitions, or input-output examples. Despite opaqueness in other parts of the AI process, these intermediates helped users reason about important analytical choices, refine their initial questions, and contribute their unique knowledge. We invite the HCI community to consider when and how intermediate artifacts should be designed to promote effective data science thinking.
\end{abstract}

\maketitle

\section{Introduction}

Generative AI systems built on large language models (LLMs) have strong potential to accelerate scientific progress. 
In data science, they can do so by lowering the barrier to accessing sophisticated techniques like data visualization, predictive modeling, and natural language processing~\cite{ren_towards_2026,rahman_llm-based_2025,mitchener_2025_kosmos}.
However, AI data science tools suffer similar pitfalls as their counterparts in AI-assisted software development: they may not actually save users time~\cite{becker_measuring_2025}, and they can introduce subtle errors that are difficult to spot but can undermine the validity of results~\cite{obrien_how_2025,nguyen_how_2024,song_evaluating_2025}.
When applying autonomous or agentic AI systems to societally-relevant data science problems, we believe that human reasoning is especially critical to success yet is currently under-supported.

\begin{figure*}
    \centering
    \includegraphics[width=\linewidth]{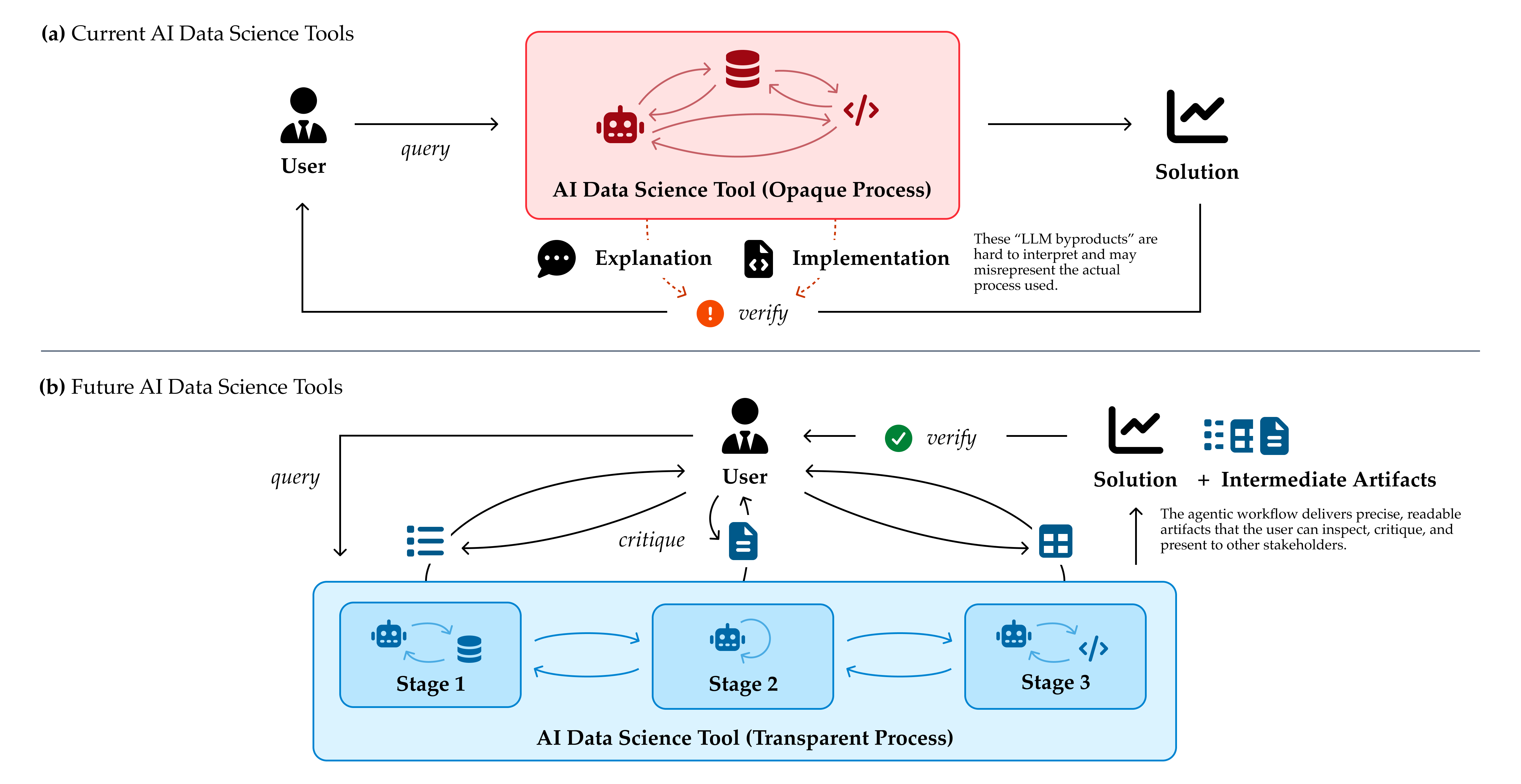}
    \caption{While current AI data science tools use complex, opaque processes that are difficult for users to verify and control (a), we propose that AI data science workflows be explicitly designed around intermediate artifacts that can foster data science reasoning and allow users to steer execution (b). How to choose the stages and intermediate artifacts to structure this transparent process is a central question for future research.}
    \label{fig:current-and-future}
\end{figure*}

To understand how modern AI data science tools may fail to fulfill their potential, let us consider Art, a (fictitious) clinician researcher who wants to understand why some patients with a certain type of cancer respond poorly to treatment.
AI-based tools such as EHR-Agent~\cite{shi_ehragent_2024} and even off-the-shelf LLMs hold the possibility to supercharge Art's data science capabilities, especially given his lack of coding expertise.
While previously he would have to find a data science expert at his institution who would be willing to work with him over a period of months, he can now ask the AI system to extract the relevant data from his institution's health record database and train models automatically.
In no more than a few weeks, Art can run a large set of models and find one with strong predictive performance.

However, there is one problem with Art's new solution: he has no idea whether his results look so good because of a real effect, or because the AI agent misinterpreted his request and solved a subtly different problem!
The system wrote hundreds of lines of code to perform its analysis, but since Art has very little data science expertise, he can't be sure whether the code is correct.
Even the short explanation that the system provided gives no indication of a potential misunderstanding that could skew the results.
As a well-intentioned researcher, Art doesn't feel comfortable publishing an analysis that he doesn't fully understand, so he spends several more weeks consulting data science experts, painstakingly validating the LLM-generated code, and re-running the tool in search of possible mistakes.

Art's predicament mirrors findings in the emerging literature on AI-assisted development~\cite{obrien_how_2025,nguyen_how_2024,becker_measuring_2025}. 
For example, one study showed that experienced developers took longer to complete tasks with AI coding tools than without them, largely due to time spent verifying outputs~\cite{becker_measuring_2025}.
Expert data analysts have begun to develop strategies for prompting generative AI so that the results are easier to verify~\cite{gu_how_2024,drosos_its_2024}, but these techniques may not be obvious to those with less data expertise.
Indeed, a study of scientists using LLMs to program found that their primary verification strategies---running code and inspecting output--failed to catch errors with scientific ramifications, and that some participants chose not to attempt to understand the generated code at all~\cite{obrien_how_2025}. 

The most obvious solution to this problem, and one that has been explored from the early days of explainable AI~\cite{lipton_mythos_2017} to modern multi-agent systems~\cite{epperson_interactive_2025,li_what_2025}, is to provide greater visibility into how the AI system arrived at its answer.
However, as AI systems for data science have become more complex and capable~\cite{liao_sagecopilot_2025,guo_ds-agent_2024,hong_data_2025,shi_ehragent_2024}, the paths they take to produce outputs have increasingly been regarded as ``means to an end.''
Moreover, LLM-generated explanations may merely increase users' reliance despite not necessarily being accurate reflections of how the answer was produced~\cite{kim_fostering_2025,obrien_how_2025}.

We argue that the problem is not an issue of transparency, but a lack of transparency designed for the user.
As illustrated in Fig. \ref{fig:current-and-future}, end-to-end systems often lack well-designed \textit{intermediate artifacts}, which we define as representations of an AI system's analytical choices that are intentionally created for human review and steering.
Unlike computational byproducts, an intermediate artifact is concise, interpretable without requiring technical expertise, and reveals underlying analytical judgments that domain experts can evaluate.
This paradigm is not new: \citeauthor{heer_agency_2019}~\cite{heer_agency_2019} described sharing editable structured representations and interactive visualizations between humans and AI to reduce effort in data analysis.
We believe this approach is more important than ever in modern data science contexts where AI systems seek to automate large parts of data analysis work, including when the problem formulation is unclear. 
Therefore, in this paper, we draw from our recent work on AI systems for data science to ask: how does the design of intermediate artifacts turn agentic AI systems from autonomous black boxes into effective tools for thought (TfT)?

\section{Case Studies}

We ground our discussion of AI data science tools in our experiences building two separate AI-powered systems in the medical domain: HACHI~\cite{feng_human-ai_2026} and Tempo~\cite{sivaraman_tempo_2025,ma_tempoql_2025}.
These works are not the only ones aiming to build more transparent AI workflows for data science~\cite{mastrianni_ai-enhanced_2025,guo_investigating_2024,schombs_conversation_2026}, but we focus on these due to our personal experiences working on their research teams.
We believe learning from the successes and limitations of these tools can help inform an HCI agenda for more autonomous future systems for data science.

\subsection{HACHI: Discovering and Annotating Concepts in Clinical Notes}

\begin{figure}
    \centering
    \includegraphics[width=\linewidth]{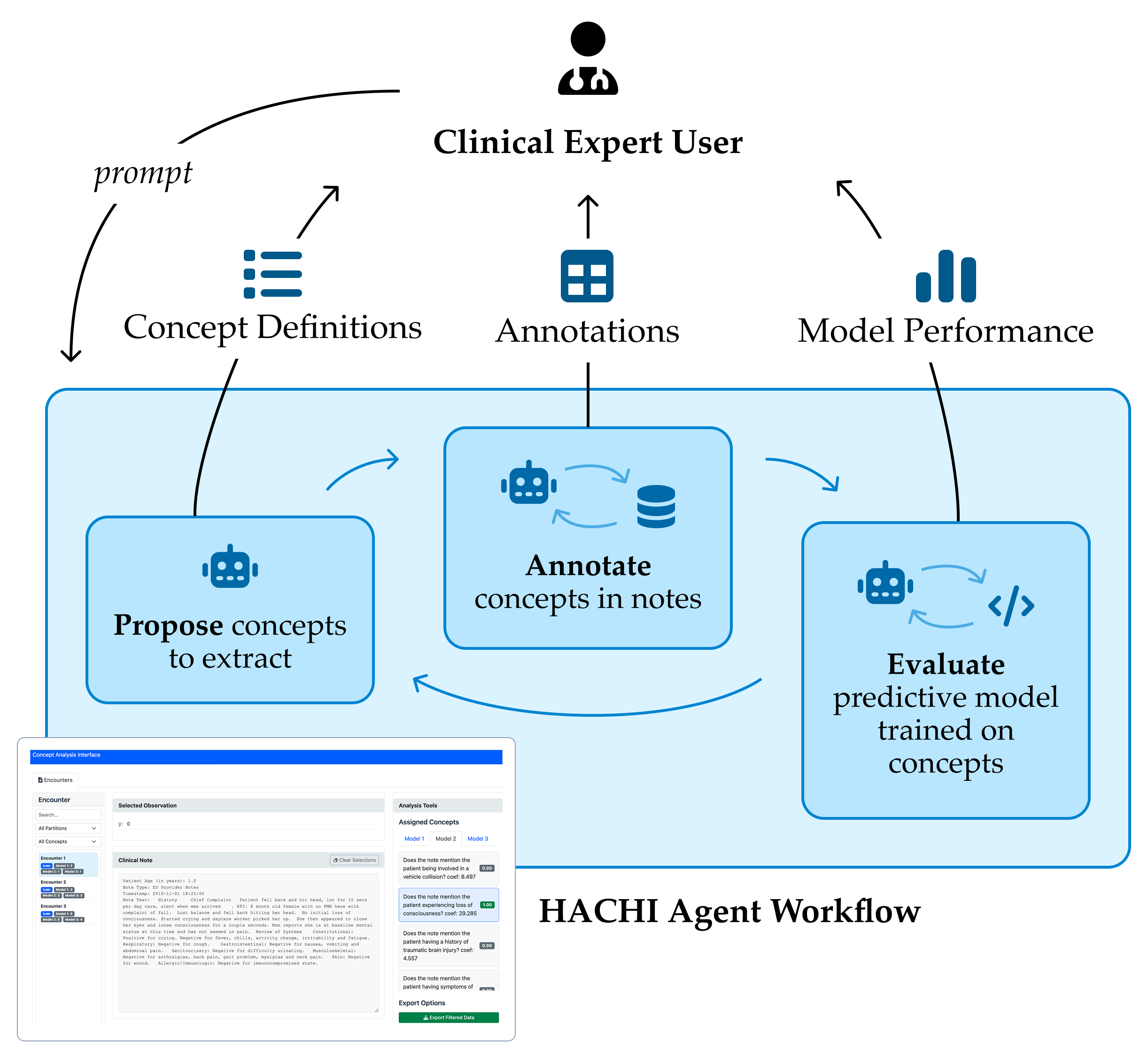}
    \caption{The HACHI workflow trains a predictive model using concepts identified from clinical notes, yielding intermediate artifacts such as the definitions of the discovered concepts, notes and their labels, and the model's performance.}
    \label{fig:hachi}
\end{figure}

The HACHI system~\cite{feng_human-ai_2026} originated as part of a collaboration with a pediatric emergency medicine researcher who was interested in building a decision support model for traumatic brain injury (TBI).
Identifying patients who should be tested for TBI is an essential task for emergency medicine providers, but the information that can predict TBI is only available in unstructured clinical notes.
The \underline{H}uman+\underline{A}gent \underline{C}o-design framework for \underline{H}ealthcare \underline{I}nstruments (HACHI), shown in Fig. \ref{fig:hachi}, was designed to support this data science challenge as well as many other similar tasks.
HACHI trains a simple statistical model on \textit{concepts} automatically annotated from clinical text; for example, a simple concept could look like ``Has the patient recently had surgery?''
The key contribution is that HACHI both identifies concepts and extracts them from notes using an LLM-powered workflow, and it surfaces these concepts to a domain expert user through an interactive interface at intermediate stages of the pipeline.
This allows users to inspect the concepts, compare them against their intuition, and reformulate the task.

Using HACHI to develop a predictive model for TBI led to several surprising findings, particularly around clinicians' involvement in the process.
In contrast to conventional workflows for building models on text data, which might obscure important nuances in code, HACHI enabled clinicians to provide feedback to the AI agent that went well beyond the developers' expectations.
For example, early on the team identified a concept (``brain bleed'') whose high predictive weight turned out to reflect data leakage---information available only after the diagnostic test the model was meant to predict---necessitating removal of contaminated cases and a restart of the pipeline.
In the next iteration, they noticed that some concepts captured note-writing style rather than patient characteristics (e.g., whether a note \textit{mentions} the Glasgow Coma Scale vs.\ whether the patient \textit{has} an abnormal score), leading them to constrain how concept definitions were phrased.
And when they discovered starkly different model performance across two hospital campuses (AUC 0.93 vs.\ 0.71), they reweighted the objective function to ensure equitable performance---a decision driven by values around algorithmic fairness that no purely automated system would have made~\cite{feng_human-ai_2026}.
Over successive rounds of this feedback, model performance and generalizability improved, with each round requiring only 1--2 hours of team review each time~\cite{feng_human-ai_2026}.
\textbf{HACHI's user interface was essential to making this feedback loop practical}: it gave collaborators full agency in deciding what needed to be reviewed and in what order, and it greatly simplified navigation over many intermediate outputs.

Nevertheless, the HACHI workflow also revealed many open questions on how best to design human-AI systems.
The data leakage issue might have been caught earlier if the manual data extraction step had itself been conducted with AI assistance, ideally supported by intermediate artifacts that made the timing of the notes visible to the whole team.
In addition, there were many other opportunities for the human team to engage with and validate intermediate outputs of the LLM pipeline.
By optimizing how often and when the human team engages with the LLM pipeline, the final results from HACHI could be even better.
For instance, HACHI currently only lets the user update the top-level prompt rather than directly editing concept definitions, but the latter may be important for providing even more precise feedback.

\subsection{Tempo: Helping Domain Experts Work with Temporal Event Data}

\begin{figure}
    \centering
    \includegraphics[width=\linewidth]{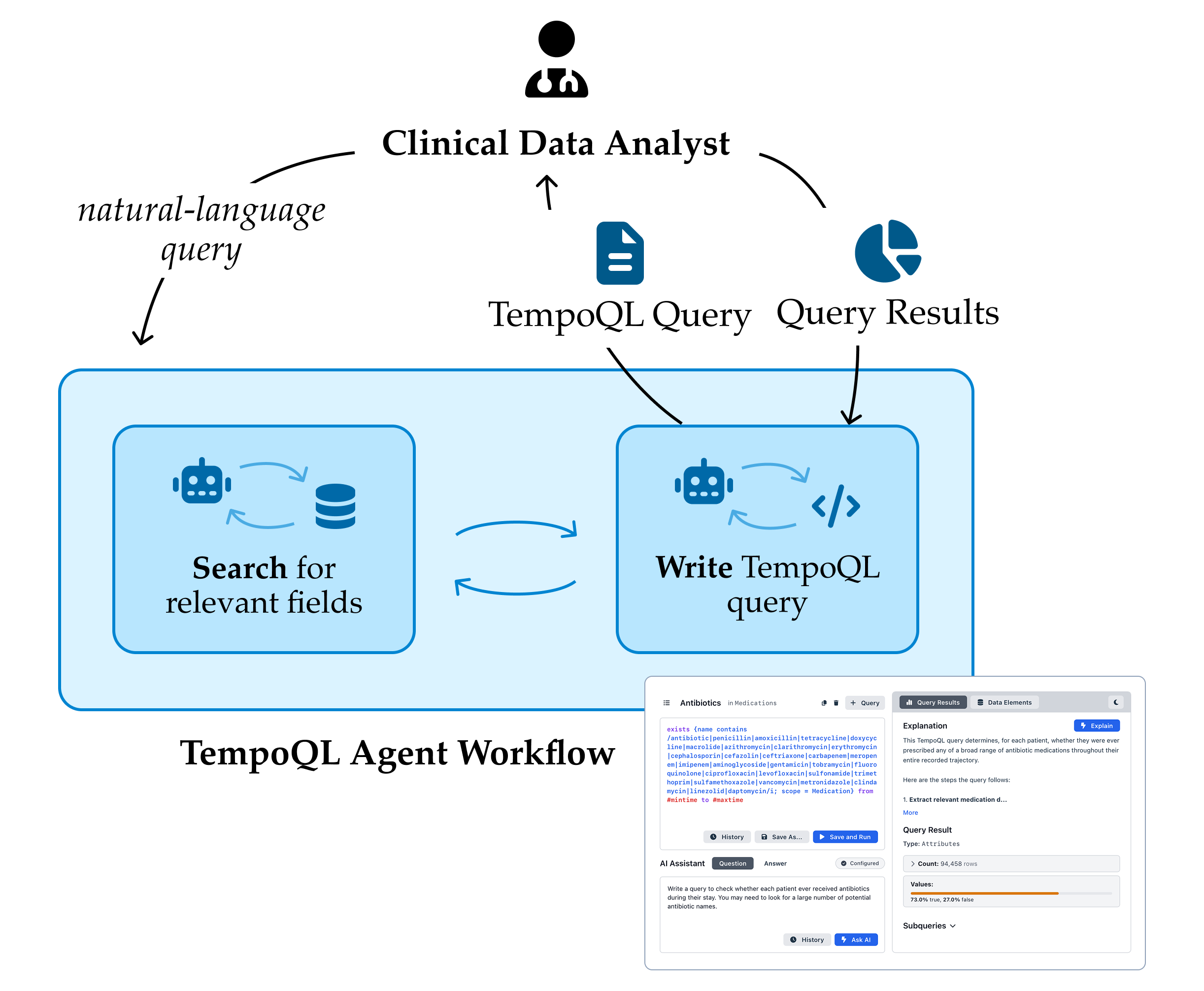}
    \caption{In Tempo, the AI Assistant translates a user's query into a readable, precise query language to extract temporal data from electronic health records. The TempoQL query and its results serve as intermediate artifacts that help the analyst decide if the data extraction was successful.}
    \label{fig:tempo}
\end{figure}

The Tempo project~\cite{sivaraman_tempo_2025,ma_tempoql_2025} began after the authors worked with several clinical research teams to build prediction models based on electronic health record (EHR) data, but found it prohibitively difficult to engage domain experts at a critical stage: defining and formatting relevant clinical events for modeling.
The goal was to design a tool for data scientists that was expressive enough to extract the EHR data they needed, while allowing domain experts to easily understand and critique how the process worked.
The final version of this system is an interactive notebook widget that allows users to write queries in a novel query language, called TempoQL, and visualize the results.
To further lower the barrier to entry for people with less data science expertise, Tempo includes an agentic AI Assistant that follows the workflow shown in Fig. \ref{fig:tempo}.
In response to a natural-language query, the AI Assistant can automatically search the EHR database for relevant concepts, then construct queries using those concepts.
Since TempoQL queries are simpler and much more concise than equivalent SQL code, clinical experts can use them to verify and edit the data extraction and aggregation procedure. 
After running the query in the interface, the user can send a summary of the results back to the AI agent to correct any errors or improve the query.

Case studies with teams of data scientists and domain experts~\cite{sivaraman_tempo_2025} showed that \textbf{the TempoQL language served as an effective intermediate for users to reason about complex data extraction workflows.}
For example, a product manager working on a web browsing tool was able to rationalize differences between queries that aggregated events at two different time intervals, and they suggested a new aggregation scheme that could combine the best of both alternatives.
We also found that off-the-shelf LLM tools, which were used to implement the AI Assistant, were 2.5x more likely to generate correct TempoQL than SQL for the same task, despite only seeing TempoQL's syntax at inference time~\cite{ma_tempoql_2025}.
These results suggest that when designing an autonomous agent for data extraction, using a simple, readable language as the intermediate artifact could have benefits for \textit{both} AI accuracy and user understanding.

Whereas HACHI showed that human involvement can help guide an AI system toward solving the right problem, Tempo illustrates how explicitly designing intermediates around human involvement can lead the AI system to produce better outputs.
Yet Tempo's intermediate query language only supports one part of the data science workflow, extraction of temporal event data.
It remains an open question how intermediate artifacts could support other tasks like exploratory analysis and predictive modeling.
Given the vastness of modern EHR datasets, Tempo also faces the potential issue of ``unknown unknowns,'' where the AI Assistant might extract only some of the relevant data fields while neglecting others.
Future designs could ask the agent to brainstorm alternatives to the solution it presents, helping users identify directions for improvement.

\section{Discussion}

Amid the proliferation of ever-more-capable AI agents in data science and other domains, it is currently unclear how (and indeed if) it is necessary to bring the human into the loop.
After all, wouldn't a complex black-box workflow that produces the correct answer save the non-data scientist user the most time?
Isn't it sufficient for the AI system developers to validate that the tool uses appropriate processes to derive its results, so end users can focus on the end product?
We argue that end-user involvement is most important when the AI needs to function as a TfT: when the problem is unclear or unspecified, when expert intuition should shape the process, and when the credibility of the output depends on the methodology.

The successes we observed in HACHI and TempoQL lead us to call on the HCI community to \textbf{design more transparent AI data science tools by intentionally scaffolding workflows with \textit{precise and steerable intermediate artifacts}}, combining \citeauthor{heer_agency_2019}'s framework with the convenience of an end-to-end AI pipeline.
For users who have insufficient expertise to think through data science problems independently, well-designed intermediate artifacts can reveal important but subtle considerations that would have required prior experience to foresee.
By pointing to ways to reformulate the task, intermediate artifacts can help bridge the ``gulf of envisioning,'' where users may not know how to specify the task they want to perform until it is at least partially completed~\cite{subramonyam_bridging_2024}.
Most importantly, they can empower non-expert users to contribute insights in the areas in which they \textit{do} have expertise.
For example, clinicians can inject their intuition about a disease into the way it is modeled or how concepts are extracted from the notes they write.

The two works discussed in this paper are a starting point for HCI researchers to explore how AI agents' intermediate artifacts could be more intentionally designed and evaluated.
We pose three open questions to guide this research agenda:

\paragraph{When and How Often Should AI Agents Surface Intermediates?}
In our view, the central question of human-centered AI agent workflow design is at which stages the agent should surface relevant artifacts for critique.
While more transparency can always be beneficial, users' time constraints often make it infeasible for them to review everything that an AI agent does~\cite{kothari_when_2025}, and inexperienced users may not always know which stages of the process are most prone to human-AI misalignment.
We suggest that intermediate artifacts should be selected by balancing (a) \textbf{how an experienced human might perform the task}, (b) \textbf{where choices depend on social values or expert intuition}, and (c) \textbf{where different answers might change the problem formulation}.
For example, TempoQL queries constitute a recipe for how to retrieve and aggregate the data, which the user can evaluate as they would a sentence in a paper's methods section.
Another potential direction for future work could be to design agentic systems that can themselves decide what user input to solicit.

\paragraph{How Can We Present Intermediate Artifacts?}
The way intermediate results are presented can directly shape how they support users' cognition.
The works discussed in this paper experimented with three modes of artifact presentation: a \textbf{precise, yet readable query language} (TempoQL), \textbf{generated natural-language prompts} (HACHI concept extraction prompts), and \textbf{input-output pairs} (HACHI concept labels).
These interface designs were effective because they did not require data science expertise to interpret and they concisely revealed key choices and misalignments.
Future work could envision alternative designs that satisfy these criteria for other stages of the data science pipeline.
For instance, a system could draw inspiration from \citeauthor{rewolinski_pcs_2025}'s Predictability-Computability-Stability (PCS) framework~\cite{rewolinski_pcs_2025} to encode best practices for data science in an AI workflow, e.g. by presenting variations of an analysis to validate its robustness.

\paragraph{How Can We Evaluate Intermediate Artifacts?}
Our case studies provided preliminary evidence that engaging domain experts through intermediate artifacts may improve the quality of their data science work. 
However, quantitatively validating this claim continues to prove challenging~\cite{heer_agency_2019}.
While prior work has examined how novices~\cite{nguyen_how_2024} and scientists~\cite{obrien_how_2025} use current LLM tools, these studies have either used simple, predefined tasks or evaluated only participants' self-reported usage.
Understanding how AI tools impact people's analyses is particularly important for high-stakes domains and open-ended problems, yet it is difficult to quantify correctness and robustness for these types of tasks.
In some ways, intermediate artifacts could make evaluation easier: for example, more edits to an intermediate output or iterations between stages could represent more effective reasoning support.
We invite the community to discuss what rigorous evaluation of AI data science workflows might look like, working towards an effective toolbox for human engagement in the era of AI agents.

\begin{acks}
Thanks to the HACHI and Tempo research teams (Avni Kothari, Ziyong Ma, Anika Vaishampayan, Richard Boyce, and others) for their contributions to the systems we discussed in this work, and to the clinicians and other domain experts whose usage of our tools and candid feedback revealed opportunities for future design. The authors gratefully acknowledge funding support from the Weill Cancer Hub West.
\end{acks}

\bibliographystyle{ACM-Reference-Format}
\bibliography{main}


\end{document}